%% file: arxiv_cardbench.tex
\documentclass{article}
\pdfoutput=1

\usepackage[nonatbib, final]{neurips_data_2024}

\usepackage[utf8]{inputenc} % allow utf-8 input
\usepackage[T1]{fontenc}    % use 8-bit T1 fonts
\usepackage{hyperref}       % hyperlinks
\usepackage{url}            % simple URL typesetting
\usepackage{booktabs}       % professional-quality tables
\usepackage{amsfonts}       % blackboard math symbols
\usepackage{nicefrac}       % compact symbols for 1/2, etc.
\usepackage{microtype}      % microtypography
\usepackage{xcolor}         % colors

\usepackage{graphicx}
\usepackage{comment}
\usepackage{soul}
\DeclareSymbolFont{extraup}{U}{zavm}{m}{n}
\DeclareMathSymbol{\varheart}{\mathalpha}{extraup}{86}
\DeclareMathSymbol{\vardiamond}{\mathalpha}{extraup}{87}
\title{CardBench: A Benchmark for Learned Cardinality Estimation in Relational Databases}

\author{Yannis Chronis 
 \And
Yawen Wang
\And
Yu Gan
\And
Sami Abu-El-Haija
\AND
Chelsea Lin
\And
Carsten Binnig
\And
Fatma \"{O}zcan
\AND
Google Inc \\
\texttt{\{chronis,yawenw,gany,haija,chelsealin,cbinnig,fozcan\}@google.com}
}

\begin{document}

\maketitle
\vspace{-0.1in}
\begin{abstract}
  Cardinality estimation is crucial for enabling high query performance in relational databases. Recently learned cardinality estimation models have been proposed to improve accuracy
  but there is no systematic benchmark or datasets which allows researchers to evaluate the progress made by new learned approaches and even systematically develop new learned approaches.
  In this paper, we are releasing a benchmark, containing thousands of queries over 20 distinct real-world databases
  for learned cardinality estimation. In contrast to other initial benchmarks, our benchmark is much more diverse and  can be used for training and testing learned models systematically. 
  Using this benchmark, we explored   whether learned cardinality estimation can be transferred to an unseen dataset in a zero-shot manner. We trained GNN-based and
  transformer-based models to study the problem in three setups: 1-) instance-based, 2-) zero-shot, and 3-) fine-tuned.
  
  Our results show that while we get promising results for zero-shot cardinality estimation on simple single table queries; as soon as we add joins, the accuracy drops. 
  However, we show that with fine-tuning, we can still utilize pre-trained models for cardinality estimation, significantly reducing training overheads compared to instance specific models.
  We are open sourcing our scripts to collect statistics, generate queries and training datasets to foster more extensive research, also from the ML community on the important problem of cardinality estimation and in particular improve on recent directions such as pre-trained cardinality estimation.
  
  %The benchmark  can be used for zero-shot, fine-tuned, or instance-based cardinality estimation learning. 
  %For this purpose, we generated training data by running  thousands of queries over 20 distinct datasets.
  %Features used in training include statistics about the data, like number of distinct values of each column in each table,
  %and query plans. The labels are the actual query result cardinalities. We generated two training datasets; one with queries over single tables with multiple filter predicates, 
  %and one with queries with a binary join between two tables, and multiple filter predicates on each table. 
   
% 1-) zero-shot (training on 19 datasets and testing on the last 20th)
 % 2-) instance-specific (training and testing one the same dataset), and 3-) fine-tuned (training on 19 datasets, and using a subset of the last
 % dataset for fine-tuning and the rest for testing).
  
  %Our results show that while we get good results for zero-shot cardinality estimation on single table queries with filter predicates; as soon as we add one join, the accuracy drops.
%  We are open sourcing our scripts to collect statistics, generate queries and final training datasets. More research is needed to handle more complex SQL queries
 % and we hope that our dataset will enable further exploration of this problem and accelerate adaption of learned cardinality models in practice.
\end{abstract}

%9 pages
\input{contents/intro.tex}  %Fatma, 1 page
\input{contents/related.tex}  %Carsten (CE Models, Benchmarks for CE), 1 page
\input{contents/corpus.tex}
%Yannis-Fatma-Sami, 2.5 pages 
\input{contents/models.tex} %Yawen-Yu, 1.5 pages
\input{contents/experiments.tex}  %Yawen-Yu, 2-3 pages
\input{contents/conclusions.tex}
% \input{contents/acks.tex}

\newpage{}
\bibliographystyle{abbrv}
\bibliography{arxiv_cardbench}  

\appendix

\input{contents/appendix.tex}

\end{document}

%% file: contents/intro.tex
\section{Introduction}
% High-level: What is CE? Does the world care about it?
% Cardinality estimation is a critical problem in modern databases for optimizing performance,
%   and has been studied extensively in database research community. 
  %With the recent advancements in ML, learned cardinality estimation has revived research in this area. So far, it has been shown that 
 % instance-based models that are trained for a specific dataset and workload outperform SOTA cardinality
  %estimation techniques significantly. Most of these models require running thousands of queries to generate 
  %training data, a very expensive training overhead. As a result, a pre-trained model that can be deployed in
  %a zero-shot manner is highly desirable. However, not much has been done on zero-shot cardinality estimation. 
  
  \textit{Cardinality estimation} (CE), the problem of determining the number of intermediate records that will be returned by a database query, is one of the key building block in modern databases for optimizing performance,
  and has been studied extensively by the database research community. 
  %With the recent advancements in ML, learned cardinality estimation has revived research in this area. So far, it has been shown that 
 % instance-based models that are trained for a specific dataset and workload outperform SOTA cardinality
  %estimation techniques significantly. Most of these models require running thousands of queries to generate 
  %training data, a very expensive training overhead. As a result, a pre-trained model that can be deployed in
  %a zero-shot manner is highly desirable. However, not much has been done on zero-shot cardinality estimation.
  CE is used by query optimizers to choose between alternative query execution plans. A query plan that produces the smallest intermediate results and processes the least amount of data usually leads to better execution times, and hence is preferred.
Thus CE is critical for optimization decisions such as choosing performant join orders, deciding on whether or not to use an index, and picking the best join method. %, to enumerate a few of its uses in query optimizers. 
  It has been shown \cite{qoleis} that "bad" cardinality estimates lead to "bad" execution plans that can have multiple orders of magnitude worse performance. CE is also critical in database recommenders to choose between alternative indexes, materializes views, and storage layouts. 
   %Query optimizer are core components of database systems that transform SQL queries into performant execution plans. Cardinality estimates are used to decide between alternative equivalent execution plan. Accurate estimate directly translate to query performance as a "bad" execution plan can have multiple orders of magnitude worse performance \cite{qoleis}. As a concrete example cardinality estimates are used to decide the order of join operators in multi-join queries. It is desirable to order perform join in the order that emits the smallest amount of intermediate results, CE is used to define that order.

%\textit{Cardinality Estimation} (CE) is a core task for database systems:
%given a database query (e.g., SQL statement) estimate the number of records that would be returned, if the query is executed. The cardinality estimates are used to choose a performant execution plan for each query amongst many alternative options. Accurate CE directly translates to improved performance and as a result has been studied extensively \cite{qoleis, harmouch2017cardinality}. 

% Database queries are executed as a graph of operators (example in Figure~\ref{fig:querygraph}), also called an execution plan. Creating an execution plan involves  defining 

% Query optimization, the process that transforms a declarative SQL query to an execution plan. Choosing a successful  Declarative SQL queries are user-friendly and agnsotic to the implementation details of a database system, because t

Traditional CE techniques used in modern database systems have well-known limitations, as they make simplistic data modeling assumptions, such as data \textit{uniformity} and \textit{independence} of columns in the tables. 
%Traditional non-learned CE techniques have well known limitations stemming from the use of simplistic data modeling assumptions (data \textit{uniformity} and \textit{independence}) and heuristics \cite{qoleis}. 
Learned CE model following an instance-based approach have been shown to improve upon the state-of-art techniques used by database systems today \cite{mscn, deepdb, balsa, DBLP:conf/sigmod/KimJSHCC22, DBLP:journals/pvldb/SunZSLT21, DBLP:journals/pvldb/NegiWKTMMKA23}. 
Despite their better accuracy, learned CE models have not been adapted in practice, due to their high training overheads, among other reasons. As a result, pre-trained CE models, which are trained on a corpus of diverse datasets on transferable features, are highly desirable to alleviate the training costs and increase adaption. 

%While there have been a lot of work on instance-based CE models, not much has been done in a zero-shot setting. 
%However deploying such models in practice is challenging as the initial model training and re-training triggered by a change in the dataset and/or workload requires can be prohibitively expensive. 
%Alternatively, a zero-shot approach where a generalizable CE model is used without the need for instance-specific training promises to overcome these limitations \cite{iris, carsten}. To achieve this zero-shot models are trained on a corpus of diverse datasets on transferable features allowing them to generalize. 

\paragraph{Our Main Contributions.}
In this paper we release \textbf{CardBench}, a benchmark containing thousands of queries on 20 distinct databases, and scripts to compute data summary statistics and generate queries. 
We generated two training datasets; one with queries over single tables with multiple filter predicates, and one with queries with a binary join between two tables, and multiple filter predicates on each table.
Further, we release a query generator and flexible infrastructure that can be used and extended to generate even more complex training queries.
In comparison to existing CE benchmarks, CardBench includes a much higher number of datasets and training data (i.e., queries and cardinalities) to experiment with and compare different types of learned CE approaches. 

To show the benchmark in-action, in this paper we have trained GNN-based and transformer-based CE models, under 3 setups: 1-) instance-based (i.e., a CE model trained only for the particular dataset), 2-) pre-trained CE model in zero-shot mode, 3-) pre-trained CE model with fine-tuning on the dataset.  
Our results show that although CE is harder to learn with a pre-trained model in a zero-shot setting, but with a small amount of samples fine-tuning a pre-trained model can achieve high accuracy as good as, if not better than the instance-based methods.
We hope that releasing CardBench to the public community will help to reduce the barrier-of-entry for experimentation with learned CE models and also foster new research, also from the ML community, and lead to further advances on this highly important problem.

%% file: contents/related.tex
\section{Background and Related Work}

% \textbf{do we need this:} We give three concrete examples, demonstrating the utility of CE. First, (i) suppose a query that reads from tables (\texttt{T1} $\leftrightarrow$ \texttt{T2} $\leftrightarrow$ \texttt{T3}), should the engine join ($\bowtie$) data as ((\texttt{T1} $\bowtie$ \texttt{T2}) $\bowtie$ \texttt{T3}), or as (\texttt{T1} $\bowtie$ (\texttt{T2} $\bowtie$ \texttt{T3}))?
% Both join orders should produce the same set of results. However, the latency of execution may differ by orders-of-magnitude.
% As a general rule-of-thumb, it is desirable to start with the join that emits fewer number of records. Here, CE can give estimates of the binary joins (\texttt{T2} $\bowtie$ \texttt{T3}) and (\texttt{T1} $\bowtie$ \texttt{T2}). (ii) Another utility of CE is to predict the type of join in distributed DB engines: \textit{distributed join} or \textit{broadcast join}?
% The latter is preferred \textbf{if} the total number of fetched bytes from one table is small-enough, so that it can be broadcasted onto (distributed) workers.
% %
% (iii) Lastly, CE is also a key problem for best physical database design; the problem of how to organize the data, which indexes and summary structures to create. 
% Wrong estimation can lead to order(s) of magnitude performance difference in query optimization.

% In the following, we discuss work related to the directions presented in this paper.

\noindent\textbf{Learned Cardinality Estimation.} Traditional  CE methods rely on heuristics and simple analytical models \cite{cesurvey} (e.g., one-dimensional histograms or sketches). Recently, there have been various approaches for learned cardinality estimation \cite{mscn, deepdb, balsa, DBLP:conf/sigmod/KimJSHCC22, DBLP:journals/pvldb/SunZSLT21, DBLP:journals/pvldb/NegiWKTMMKA23,DBLP:conf/icde/NegiMMTKA20,DBLP:journals/pvldb/ZhuWHZPQZC21,DBLP:journals/pvldb/WangCLL21}.
While early learned approaches used classical regression models, more recently deep learning based approaches \cite{mscn,cebench2} have been proposed which show highly promising accuracy over traditional approaches.
%While many approaches with different model architectures \cite{DBLP:journals/pvldb/NegiWKTMMKA23,DBLP:conf/icde/NegiMMTKA20,DBLP:journals/pvldb/ZhuWHZPQZC21,DBLP:journals/pvldb/WangCLL21} have been suggested for deep learning based cardinality estimation, what many of these approaches have in common is the use of features such as table and attribute names of a given query plan, as input to predict the cardinalities for a query.
All of these methods are instance-based: they utilize instances-specific features like names of tables and columns; they are evaluated and trained on the same database. 
There are two main approaches: workload driven and data driven. Workload driven approaches need to run a representative set of queries (10 to 100 thousands) to collect true cardinalities as labels for the training samples, incurring a very high overhead. These models also need to be re-trained as the representative workloads shift.
%As such, these models are instance-specific since they need to learn a cardinality estimator in particular i.e., for one dataset which involves a representative set of query plans that need to be executed to collect the true cardinalities which can be used for training such models.
%However, this causes extremely high overheads for training data collection since it involves executing 10 to 100 thousand of query plans per dataset. 
To avoid this high overhead of running thousands of queries to collect the true cardinalities per dataset, more recently so called data-driven learned cardinality estimators have been proposed \cite{deepdb,neurocard}.
In contrast to the workload-driven approaches,  data-driven approaches learn the multi-variate data distribution within and across tables without running any query plan but simply by learning data distributions of the base tables.
%That way, the overhead for training can be significantly reduced.
Data driven approaches still require one model per dataset and need to be re-trained as the data gets updates, which still incurs training overheads.
%needs to be trained per dataset and must be kept up-to-date the data distributions change, which causes training overhead.
As such, very recently some first generalizable cardinality estimators \cite{iris} have been proposed that can be used on  datasets in a zero-shot manner; i.e., the model can be used without the need to learn one instance-specific model per dataset. Moreover, zero-shot models also promise that they can automatically provide accurate estimates even if the underlying data of the dataset is updated (i.e., new rows are added or existing ones are updated or deleted). 
The core idea is that such zero-shot models are pre-trained on a broad spectrum of different datasets and use transferable features, such as table sizes,  allowing them to generalize to unseen datasets.
%However, currently existing generalizable zero-shot models \cite{iris} are limited to simple queries which involve only one table at the moment.
We argue that more research is needed on pre-trained models on more compelx queries. For this purpose, we provide a very first large-scale training corpus to foster research in this direction.
%on zero-shot cardinality estimation on more complex queries.

\paragraph{Cardinality Estimation Benchmarks.} To  benchmark cardinality estimation models, the early papers have mainly relied on either home-grown benchmarks with non-public datasets or they have used some available benchmarks that have been originally designed for other purposes. One central benchmark that has been used so far extensively by many papers is the Join Order Benchmark (JOB) \cite{job}, which is based on real-world data from IMDb, containing interesting data distributions and correlations. JOB contains 79 queries, which challenge cardinality estimation. This is in stark contrast to many other database benchmarks such as the TPC benchmarks which use synthetic data that is way simpler and thus does not represent a real challenge for cardinality estimation.
However, one benchmark with one dataset is not sufficient to benchmark learned cardinality estimation and show its robustness towards various different real-world domains. As such, in the recent years, there have been further efforts to create additional benchmarks that can be used to test cardinality estimation models. For example, \cite{cebench1} proposes a new benchmark, which contains one new real-world dataset STATS and a generated query workload that is more diverse in terms of query and data characteristics than JOB.
Another benchmark \cite{cebench2}, which aims in a similar direction uses two different datasets - IMDB and StackExchange (SE) - and contains  16K unique queries and true cardinalities.
However, these two benchmarks still mainly target instance-specific models which can be seen by the fact that they only contain one or two different datasets, which is clearly no sufficient for testing pre-trained  models that require a representative set of different datasets.
%which allows to pre-train and test models on disjoint data.
In this paper, we thus present a benchmark that targets the pre-trained cardinality estimation and thus includes a much more diverse set of 20 different datasets which can be used for pre-training and testing.

%\paragraph{Zero-shot Models for Databases.}
%In addition to cardinality estimation, there exists other approaches that use zero-shot learning for databases.
%One of the initial work by Hilprecht \& Binnig \cite{carsten} which proposes zero-shot models for estimating \textit{query execution time}. They train a model on 19 workloads and test it on an unseen 20th workload.
%Our contribution allows similar models, on cardinality (rather than runtime) estimation.

%% file: contents/corpus.tex
\section{CardBench } 
\label{sec:benchmark}

This section describes \textbf{CardBench}, an open-source benchmark\footnote{All data are released under the CC BY 4.0 International license, all source files are released under the Apache 2.0 license. \url{https://github.com/google-research/google-research/tree/master/CardBench_zero_shot_cardinality_training}} for CE, 
%that contains 20 distinct datasets and thousands of queries of different complexities. Our goal is to foster further research in the area of learned CE, 
with a focus on enabling the training and testing of pre-trained zero-shot CE models but also instance-specific CE as well as fine-tuning pre-trained CE models. 
%Further, we release a query generator and flexible infrastructure that can be used and extended to generate complex training queries, for the existing or new datasets. We also release the calculated statistics and two complete training datasets as re-creating them can be time and resource intensive.

% In addition to datasets and queries we also release the training data for two workloads (Table~\ref{table:databases}) as calculating statistics, running queries and generating training data can be extremely time and resource consuming. 

% and the pipeline of scripts used to create queries and training data are releases as open source The scripts that perform statistics calculation, the query generator and create training data are designed to be extensible.
% CardBench is made up of datasets and their statistics, queries and trainind data.

% Sections~\ref{sec:bench-datasets-selection}-\ref{sec:bench-queries-training-data} describe the design of the benchmark, Section~\ref{sec:bench-pipeline} describes how the benchmark was created, Section~\ref{sec:bench-release} summarizes the benchmark and set of scripts

% Specifically we release two training datasets that are made up of thousands of queries. The first trainind dataset contains queries over single tables with multiple filter predicates, and the other one that contains a binary join between two tables, and multiple filter predicates on each table. 

\noindent\textbf{Dataset Selection:} % \label{sec:bench-datasets-selection}
Table~\ref{table:databases} list the CardBench datasets. The datasets were chosen to be diverse, complex and cover a wide range of data distributions to stress CE models. All the datasets are publicly available or are based on publicly available datasets (we list the sources in the CardBench repository). Datasets marked with ($\sim$) in Table~\ref{table:databases} are created by random sampling the original datasets to reduce their size. Smaller datasets reduce the cost of statistics calculation and running queries. For example, the size of the original \emph{github\_repos} dataset is 3 petabytes, using random sampling the version we use in CardBench is 300 gigabytes in size. The number of datasets included with CardBench was chosen to allow for training zero-shot models, which is validated by our experiments and previous work~\cite{carsten},

\noindent\textbf{Queries and Training Data:} %\label{sec:bench-queries-training-data}
Our training data are SQL queries represented as graphs annotated with dataset statistics and the query cardinality as the label (Table~\ref{table:features}, right). CardBench includes two sets of training data, \emph{Single Table} and \emph{Binary Join} (Table~\ref{table:databases}). \emph{Single Table} queries filter a single table using 1 to 4 filter predicates. \emph{Binary Join} queries join two tables that are also filtered with 1 to 3 filter predicates per table. The two training dataset configurations can demonstrate the challenges of more complex queries on CE models (see Section \ref{sec:experiments}). To obtain the training labels (i.e. the query cardinalities) we used Google Big Query, executing the queries of \emph{Single Table} and \emph{Binary Join} required 7 cpu time years.
\textbf{As such, releasing the queries with their true cardinalities significantly lowers the entry barrier for research on learned CE models.}

% The table without Dataset ID. 
\begin{table}
\small
\caption{Dataset and Workload Information}
\label{table:databases}
\begin{center}
\small
\begin{tabular}{r  c  c  c }
\toprule
\textbf{Database Name} & \textbf{Table Count} & \textbf{Single Table Queries} & \textbf{Binary Join Queries} \\
\midrule
% tpch\_10G & 8 & 11,727 & 13,181 \\
% accidents & 3 & 9,133 & 8,547 \\
% airline & 19 & 6,625 & 13,114 \\
% $^{(\sim)}$ cms\_synthetic\_patient\_data\_omop & 24 &  12,403 & 7,877 \\
% $^{(\sim)}$ covid19\_weathersource\_com & 4 & 9,370  & 10,186  \\
% $^{(\sim)}$ ethereum\_blockchain & 7 & 19,759 & 17,424  \\
% $^{(\sim)}$ geo\_openstreetmap & 16 & 16,532  & 15,000  \\
% $^{(\sim)}$ github\_repos & 8 & 5,272 & 6,374 \\
% $^{(\sim)}$ human\_variant\_annotation & 26 & 13,888  & 14,085  \\
% $^{(\sim)}$ idc\_v10 &  19 & 8,657 & 6,505  \\
% $^{(\sim)}$ open\_targets\_genetics &  13 & 9,774  & 9,607  \\
% $^{(\sim)}$ samples & 8 & 8,899 & 11,369  \\
% $^{(\sim)}$ stackoverflow &14 &  14,314  & 12,773  \\
% $^{(\sim)}$ usfs\_fia & 11 & 13,925 & 11,292  \\
% $^{(\sim)}$ uspto\_oce\_claims & 10 &  5,935  & 6,123 \\
% $^{(\sim)}$ wikipedia & 15 & 6,610 & 6,020 \\
% $^{(\sim)}$ crypto\_bitcoin\_cash & 2 & 14,317 & 12,404 \\
% consumer & 3 & 5,962  & 5,571  \\
% movielens & 13 & 14,497 & 12,524 \\
% employee & 6 & 12,681 & 10,417 \\
accidents	&	3	&	9125	&	8454	\\
airline	&	19	&	6568	&	13096	\\
$^{(\sim)}$ cms\_synthetic\_patient\_data\_omop	&	24	&	12229	&	7833	\\
consumer	&	3	&	5961	&	5571	\\
$^{(\sim)}$ covid19\_weathersource\_com	&	4	&	9366	&	10186	\\
$^{(\sim)}$ crypto\_bitcoin\_cash	&	2	&	14315	&	12404	\\
employee	&	6	&	12675	&	10417	\\
$^{(\sim)}$ ethereum\_blockchain	&	7	&	19756	&	17424	\\
$^{(\sim)}$ geo\_openstreetmap	&	16	&	16346	&	15000	\\
$^{(\sim)}$ github\_repos	&	8	&	5268	&	6789	\\
$^{(\sim)}$ human\_variant\_annotation	&	26	&	13862	&	14085	\\
$^{(\sim)}$ idc\_v10	&	19	&	8613	&	6470	\\
movielens	&	13	&	14488	&	15757	\\
$^{(\sim)}$ open\_targets\_genetics	&	13	&	9734	&	9058	\\
$^{(\sim)}$ samples	&	8	&	8893	&	11369	\\
$^{(\sim)}$ stackoverflow	&	14	&	14305	&	12773	\\
tpch\_10G	&	8	&	11727	&	13181	\\
$^{(\sim)}$ usfs\_fia	&	11	&	13822	&	11695	\\
$^{(\sim)}$ uspto\_oce\_claims	&	10	&	5925	&	6123	\\
$^{(\sim)}$ wikipedia	&	15	&	6659	&	7375	\\
\bottomrule
\end{tabular}
\end{center}
\end{table}

The number of queries we include per dataset varies, as shown in Table~\ref{table:databases}. The query generator (described in Section \ref{sec:sqlgeneration}) uses dataset information and statistics and follows a random process to create realistic queries. Although we generate the same number of queries per datasets, during query execution we filter out queries that are duplicate, return zero results or timeout while running. As a result, the successfully run queries that become the training data vary per dataset. We opted not to cap the number of queries per dataset and leave it to the benchmark user to use as many queries as they deem useful. For example, for the experiments shown in Section \ref{sec:experiments} we use the same number of queries across all datasets to ensure a fair cross-comparison.

\noindent\textbf{Accuracy Metric} %\label{sec:bench-accuracy-metric}
The \textit{q-error} is a common metric in database systems research for evaluating the accuracy of CE models~\cite{cebench1,cesurvey,iris}. It calculates the relative deviation of the predicted cardinality from the true cardinality for a given query. The Q-error for a single query is calculated as follows: 
% \vspace{-0.1in}
\begin{equation}
    \textrm{Q-error} = \max \big( \frac{\hat{y}}{y}, \frac{y}{\hat{y}} \big) \in [1, +\infty),
\end{equation}

where $y$ and $\hat{y}$ represent true and predicted cardinalities, respectively. We particularly focus on P50 and P95 q-errors aggregated across all queries within a dataset to evaluate the models' accuracy in common and tail cases. 
We refer to Section~\ref{sec:experiments.model_configs} for the details of how CardBench is used to train and evaluate zero-shot, instance and fine-tuned CE models.

\begin{figure}[t]
	\centering
	 \includegraphics[width=14cm,height=1.2cm]{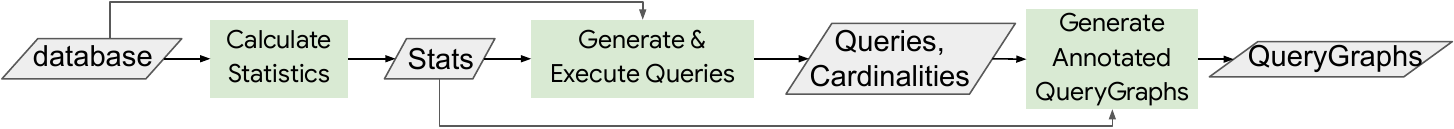}
	\caption{CardBench Benchmark Creation Steps. The scripts to create the benchmark along with the training data (query graphs) released as open-source.}
	\label{fig:pipeline}
	\vspace{-1.5ex}
\end{figure}

\subsection{Benchmark Creation}
\label{sec:bench-pipeline}

The process we follow to create CardBench is illustrated in Figure \ref{fig:pipeline} and comprises of three major steps a) statistics calculation, b) query generation and execution, c) annotated query graph generation. The first and second steps of the process involve running SQL queries to calculate statistics and collect the query plans and training labels. Thus, creating CardBench is expensive and for that reason alongside the pipeline that creates the benchmark we release the statistics of the datasets, the queries with their cardinalities and the query graphs.

% Our goal is for this benchmark to be database agnostic and to contain diverse datasets that stress the abilities CE models and will allow the model to generalize across datasets and workloads. 

% CE models need the actual result size (i.e. query cardinalities) as labels, require running the training queries to observe their result size.

\subsubsection{Dataset Statistics Calculation}
\label{sec:stats-calculation}
%We focus on transferable features and not the features such as table and attribute names specific to a dataset or workloads. This is similar to non-learned CE methods and different from the approach of existing instance-based learned CE models. 
In CardBench we include a set of data statistics and summaries which can be used as input features for CE models, we expect traditional query optimizers to calculate and store most of the statistics we use, but as not all query optimizers maintain the same set of statistics or calculate them in the same way, the first step in our infrastructure is statistics calculation. Therefore the statistics used by CardBench are database system agnostic. The calculated statistics are stored in a database and used to create the SQL queries as well as the training data.
% As this step can be expensive we also release the result of the statistics calculation. 
Table~\ref{table:features} lists the statistics we calculate/collect.

\noindent\textbf{Primary foreign key relationships} (pk-fk) are part of a database schema design~\cite{cow}. A pk-fk relationship is a logical connection between the columns of two tables semantically linking their information. Naturally we want to use pk-fk relationships when creating synthetic workloads. A challenge we faced is that open source datasets do not usually specify pk-fk relationships, instead we discover pk-fk relationship between columns of different tables using the steps outlined in Appendix~\ref{sec:pkfk}.

\begin{table}[h]
\caption{\textit{(left)} Query Graph (heterogeneous) features.
Features can be numeric ($_{\in \mathbb{R}}$) or categorical ($_{\in \mathbb{Z}}$).
Symbol $^\vardiamond$ denote features utilized in query workload generation, and $^\varheart$ denotes ones used by our models (\S\ref{sec:experiments}).
Notes:
$^\textrm{(i)}$ \S\ref{sec:stats-calculation}; 
$^\textrm{ (ii)}$ Always = ``Pearson'';
$^\textrm{ (iii)}$ $\in \{\texttt{join}, \texttt{scan} \}$;
$^\textrm{ (iv)}$ $\in \{ =, >, <, \ge, \le, \texttt{IS} , \texttt{IS NOT}, \texttt{BETWEEN}, \texttt{AND} \}$. \textit{(right)} Query Graph depiction with corresponding SQL query statement.
%
%Column features lists the features (or statistics) we calculate in and use as input in the Workload Generation (column Workload) and/or to generate training data (Training column). Column Node Type groups column Feature. For example, feature rows refers to the number of rows of a table, is of type Numerical and is used to generate workloads and as a training data feature. Type N is Numerical, C is Categorical
}
\label{table:features}
\definecolor{mygreen}{rgb}{0.86, 0.91, 0.84}
\definecolor{mypink}{rgb}{0.93, 0.80, 0.80}
\definecolor{myyellow}{rgb}{0.99, 0.95, 0.82}
\definecolor{myblue}{rgb}{0.83, 0.88, 0.95}
\definecolor{myred}{rgb}{0.92, 0.32, 0.29}
%\centering
\begin{minipage}{0.71\linewidth}
\small
\begingroup
\setlength{\tabcolsep}{2pt} % Default value: 6pt
\begin{tabular}{ c l c c l } % l l c c
%\toprule
\cmidrule[1pt](rr){1-2}\cmidrule[1pt](rr){4-5}
\textbf{Node Type} & \textbf{Feature} & & \textbf{Node Type} & \textbf{Feature}  \\ %  & \textbf{Description} & \textbf{Type } 
% & \textbf{Workload} & \textbf{Training} \\ 
\cmidrule(lr){1-2}\cmidrule(lr){4-5}
tables & rows$_{\in\mathbb{R}}^{\vardiamond\varheart}$  % & Number of rows & N &  Yes & Yes \\
% attribute
& & attributes  & name$_{\in\mathbb{Z}}^{\vardiamond}$\\
\fcolorbox{black}{myblue}{\rule{0pt}{3pt}\rule{3pt}{0pt}}       & name$_{\in\mathbb{Z}}^{\vardiamond}$ % & Table name& C &  Yes & No \\ 
% attribute
 &  &  \fcolorbox{black}{mygreen}{\rule{0pt}{3pt}\rule{3pt}{0pt}}  & data\_type$_{\in\mathbb{Z}}^{\varheart}$ \\ %               & Attribute type& C & No & No \\
% \cmidrule(lr){1-2}
% dataset & pk-fk realtionships$_{\in\mathbb{Z}}^{\vardiamond\textrm{ (i)}}$
% % attribute
% & &             & is\_clustering$_{\in\mathbb{Z}}$ \\
\cmidrule(lr){1-2}
correlations & correlation$_{\in\mathbb{R}}^{\varheart}$ 
% attribute
& &             & null\_frac$_{\in\mathbb{R}}^{\vardiamond\varheart}$   \\
\fcolorbox{myred}{white}{\rule{0pt}{3pt}\rule{3pt}{0pt}}             & validity$_{\in\mathbb{Z}}^{\varheart}$ 
% attribute
& &             & num\_unique$_{\in\mathbb{R}}^{\varheart}$ \\
             & type$_{\in\mathbb{Z}}^\textrm{ (ii)}$  
% attribute
& &             & percentiles\_100\_numeric$_{\in\mathbb{R}}^{\vardiamond\varheart}$ \\
\cmidrule(lr){1-2}
ops \fcolorbox{black}{mypink}{\rule{0pt}{3pt}\rule{3pt}{0pt}} & operator$_{\in\mathbb{Z}}^{\varheart\textrm{ (iii)}}$ 
% attribute
& &             & percentiles\_100\_string$_{\in\mathbb{Z}}^{\vardiamond}$   \\
\cmidrule(lr){1-2}
predicates & predicate\_operator$_{\in\mathbb{Z}}^{\varheart\textrm{ (iv)}}$
% attribute
& &             & min\_string$_{\in\mathbb{Z}}^{\vardiamond}$  \\
\fcolorbox{black}{myyellow}{\rule{0pt}{3pt}\rule{3pt}{0pt}}           & estimated\_selectivity$_{\in\mathbb{R}}^{\varheart}$  
% attribute
& &             & max\_string$_{\in\mathbb{Z}}^{\vardiamond}$  \\
           & offset$_{\in\mathbb{R}}^{\varheart}$  
% attribute
& &             & min\_numeric$_{\in\mathbb{Z}}^{\vardiamond}$   \\
           & constant$_{\in\mathbb{Z}}$  
% attribute
& &             & max\_numeric$_{\in\mathbb{Z}}^{\vardiamond}$  \\
           & encoded\_constant$_{\in\mathbb{R}}$  
% attribute
& &             &    \\
% & &  &  &    \\
% min\_str$_{\in\mathbb{Z}}^{\vardiamond}$
% max\_str$_{\in\mathbb{Z}}^{\vardiamond}$
% is\_partitioning$_{\in\mathbb{Z}}^{\varheart}$
\cmidrule[1pt](lr){1-2}\cmidrule[1pt](lr){4-5}
\end{tabular}
\endgroup
\end{minipage}
\begin{minipage}{0.28\linewidth}
\begin{tabular}{|p{0.95\linewidth}|}
\hline
\includegraphics[width=0.95\linewidth]{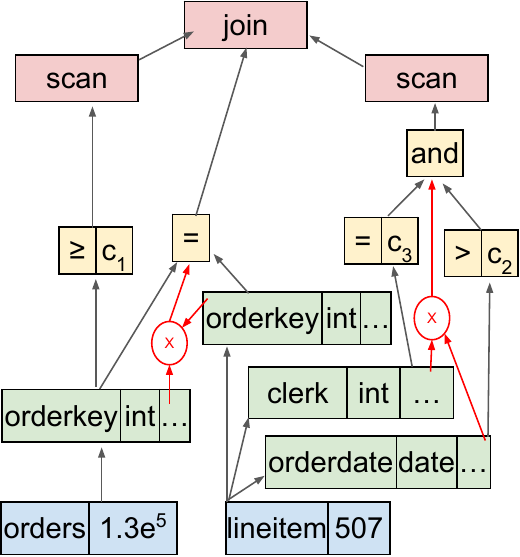} \\
\hline
\scriptsize
\texttt{SELECT * 
FROM \sethlcolor{myblue}\hl{orders o}, \hl{lineitem l}
\newline WHERE \sethlcolor{mygreen}\hl{o.orderkey} \sethlcolor{myyellow}\hl{=} \sethlcolor{mygreen}\hl{l.orderkey}
\newline \sethlcolor{mypink}{AND} \sethlcolor{mygreen}\hl{o.orderkey} \sethlcolor{myyellow}\hl{>= c1}
\newline \sethlcolor{mypink}{AND} \sethlcolor{mygreen}\hl{l.orderdate} \sethlcolor{myyellow}\hl{> c2} 
\newline \sethlcolor{myyellow}\hl{AND} \sethlcolor{mygreen}\hl{l.clerk} \sethlcolor{myyellow}\hl{= c3}
\newline \sethlcolor{mypink}{AND} \sethlcolor{mygreen}\hl{o.orderkey} \sethlcolor{myyellow}\hl{=} \sethlcolor{mygreen}\hl{l.orderkey}
} \\
\hline
\end{tabular}
\vspace{-6ex}
\end{minipage}
\end{table}

% \begin{table}
% \caption{Calculate Dataset Statistics YC:consider merging with table 3 to save space and redundant info}
% \label{table:stats}
% \begin{center}
% \begin{tabular}{c | c }
% \toprule
% Statistic & Description\\
% \cmidrule(lr){1-2}
% row count & per table \\
% number of unique values & per column \\
% NULL fraction & The percentage of NULL values per column \\
% min/max values & per column \\
% mean value & per column when datatype is numeric \\
% unique values & per column when datatype is categorical (and num unique values < 10k) \\
% histogram & size: 100 buckets, collected \\
% random table samples & 4'000 row sample per table\\
% Pearson Correlation & calculated on the 4'000 row \ samples between numeric columns of the same table\footnote{} \\
% frequent words & for string columns (max: 10k words, each appears more than 1\%) \\
% PK-FK relationships & \\
% \bottomrule
% \end{tabular}
% \end{center}
% \end{table}

\subsubsection{SQL Query Generation and Execution}
\label{sec:sqlgeneration}
This step of the process generates a set of SQL queries per dataset and executes it to collect the query cardinalities, the training labels. We use the workload generator proposed by ~\cite{carsten}~\footnote{https://github.com/DataManagementLab/zero-shot-cost-estimation, license: Apache License 2.0}. The generator takes as input the calculated statistics for each dataset and also a configuration that specifies the SQL query shapes to be produced. We have modified the original workload generator so that all queries return the cardinality as their final result. 

\subsubsection{Annotated Query Graph Generation}
\label{sec:trainingdata}
After the queries are executed, we convert the SQL queries into a graph representation which is annotated with statistics.
We focus on transferable features and not dataset or workload specific features such as table and attribute names. This approach enables learning of zero-shot CE models. %This is similar to non-learned CE methods and different from the approach of existing instance-based learned CE models. 
An example is shown in Table~\ref{table:features}, on the right we present the sql query and its representation as a graph. The graph nodes are annotated with the features listed on the left of Table~\ref{table:features}. The query graph has table nodes (in blue), column nodes (in green), operator nodes (in red), predicates (in yellow) and correlations across column of the same table (white circles with red outline). The only feature that is calculated as part of this step is the predicate \emph{estimated\_selectivity}, as both dataset statistics and query are needed. Selectivity is a fraction that represents the part of the input that satisfies a predicate \cite{cow} and is used to estimate cardinality. We estimate selectivity for filter predicate nodes using the methods described in ~\cite{accesspath}.

%% file: contents/models.tex
\section{Cardinality Estimation Models} %Yawen-Yu, 1.5 pages 
\label{sec:models}

In this section, we discuss details of our model for which supports zero-shot CE.
% , we show the results of using our benchmark on these models. 
\vspace{-0.1in}

\subsection{Data Engineering}

\textbf{Features}: The set of features (\textit{numerical} or  \textit{categorical}) that the models use are specified in Table~\ref{table:features}. In order to train a zero-shot model that can generalize to new datasets, we only include transferable features that are dataset-agnostic. This means that we exclude dataset-specific information such as table or column names. Furthermore, we select detailed statistics (e.g., histogram, correlation) as input features so the model can learn the underlying data distribution.
% from different datasets.  

\textbf{Pre-processing}: For numerical features and the cardinality label, we employ a data transformation process to enhance their representation. First, we apply a logarithmic transformation to the data, which helps to reduce the impact of outliers. We then perform standardization to ensure that all features are on the same scale. For categorical features, we adopt one-hot encoding to allow effective learning of relationships between different categories. Additionally, we filter out queries with zero cardinality or zero predicates. In practice, such queries are relatively rare, and their inclusion could introduce noise into the training process. 
 
\vspace{-0.1in}

\subsection{Graph Neural Network Model}
The graph representation of SQL queries provides a natural way to capture the relationship between different elements of a query. This representation allows us to directly leverage Graph Neural Network (GNN) which is designed to learn from graph-structured data. We build a GNN model to predict cardinality of queries based on the design of the GNN-based cost model from \cite{carsten}. The GNN model first initializes the hidden state for each node by applying a node-type specific Multi-Layer-Perceptrons (MLP) to the node’s input feature vector (a concatenation of the input feature values of that node). It then updates the node hidden states via bottom-up message passing from leaf to root nodes, following the natural query processing order. Finally, the hidden state of the root node is read out as the learned graph-level embedding. This embedding is passed into an MLP-based cardinality model to get the final prediction.

\subsection{Graph Transformer Model}
In addition to GNN, Graph transformer is an emerging type of neural network architecture specifically designed for graphs. Similar to traditional transformers, which capture the relationship between tokens in a sequence with self-attention, graph transformer extends the attention mechanism to graphical data, where the relationships between different nodes are constrained by the graph structure~\cite{yun2019graph}. 
% TODO(gany@): draw graph_transformer_overview diagram (https://docs.google.com/drawings/d/1YyIRTEr5vP2ZqD_Lzff6nGkzDTQAJfdZup41UB4ye0Y/edit?resourcekey=0-tiRwHpOsJJ2rE2eoWXPB6Q)
Compared to traditional transformers, we make following adaptation\footnote{Figure \ref{fig:graph_transformer_overview} in the appendix illustrates the architecture of the attention block of the graph transformer model designed to embed SQL query graphs. }: 

\textbf{Heterogeneous input embedding layer}: The SQL query graph is a heterogeneous graph with multiple node types that contain different features. For each node type, we use a separate input embedding layer to project node features to a fixed-dimension input node embedding. 

\textbf{Shortest distance spatial encoding}: Following the methodology of Graphormer~\cite{ying2021transformers}, we compute the shortest path distance between node pairs to construct a spatial encoding matrix. This matrix is incorporated as a learnable bias that adds to the input before the softmax in the self-attention block.

\textbf{Directional causal mask}: To ensure information flow adheres strictly to the DAG's structure, we implement a causal masking mechanism based on the topological ordering of the nodes.

\textbf{Virtual node readout}: Inspired by Graphormer, we augment the query graph by introducing a virtual node as the direct child of all other nodes. Given that the virtual node attends to all other nodes in the query graph, its embedding serves as the graph-level representation for cardinality prediction.

%% file: contents/experiments.tex
\vspace{-0.1in}
\section{Experimental Findings}
\label{sec:experiments}

In this section, we present our empirical findings, evaluating the accuracy of various model configurations for cardinality prediction across all datasets in CardBench.
\vspace{-0.1in}

\subsection{Evaluation Setup}

\subsubsection{Model Configurations}
\label{sec:experiments.model_configs}
We assess cardinality prediction performance using three distinct ML model configurations: instance-based, zero-shot, and fine-tuned. 

\textbf{Instance-based Models: } In this configuration, we train and evaluate individual models using a single dataset. For each dataset, queries are randomly partitioned into training and validation sets with 85:15 train-validation split. An additional 500 queries are reserved as a standalone test set to evaluate final model accuracy.

\textbf{Zero-shot Models: } This configuration investigates the generalizability of cardinality prediction models to out-of-distribution data, i.e., data from another dataset apart from the training data. We train and validate the model on queries from 19 datasets, maintaining the 85:15 train-validation split. The model is then tested on the standalone test set of the remaining 20th dataset. This test set contains the same queries used in the instance-based setting to ensure a fair comparison.

\textbf{Fine-tuned Models: } This configuration investigates whether fine-tuning a pre-trained zero-shot model improves accuracy and sample efficiency.  We fine-tune the model initially trained on 19 datasets using the 20th dataset. The fine-tuned model's accuracy is then evaluated using the same 500 holdout queries from this 20th dataset.

The average training times under each configuration are\footnote{Listed in instance-based, zero-shot, fine-tuned order.} 1.3hr, 1.4hr and 11min (single-table queries), and 1.3hr, 1.5hr, and 11min (binary join queries) for the GNN-based model on a 8-core VM (4GB memory). For the Transformer-based model the average training times are  3.3hr, 11.8hr and 1.6hr (single-table queries), 11.1hr, 11.1hr and 2.1hr (binary join queries) on a V100 GPU. %The zero-shot training time is not significantly higher than instance-based despite having much more training data because we cap its maximum number of training epochs to 20. 
We observe that training the zero-shot model over extended epochs can often result in lower accuracy due to over-fitting to the training set. Therefore, we cap the maximum number of training epochs for zero-shot training to 20, and set a maximum training epochs of 100 for the other two configurations. 

\vspace{-0.08in}
\subsubsection{Baseline} 
We compare the learned cardinality approaches against a baseline algorithm that represents heuristics employed in conventional DBMS such as PostgreSQL\footnote{https://www.postgresql.org/}. This baseline algorithm assumes data independence between columns and computes the selectivity of conjunctive (AND) 
% and disjunctive (OR)
predicates using the formula: $sel(p_1 \land p_2) = sel(p_1) \cdot sel(p_2)$.
% and $sel(p_1 \lor p_2) = sel(p_1) + sel(p_2) - sel(p_1) \cdot sel(p_2)$.
The selectivity of other relational predicates is estimated according to the description in Section \ref{sec:trainingdata}. 

For binary joins, the baseline approach makes the simplifying assumption that the smaller table contains the primary key, which participates in the join condition, hence the cardinality of the binary join is computed as the estimated output cardinality of the scan operator of the larger table (an assumption commonly made by database systems).   

% \subsubsection{Evaluation Metrics}
% The \textit{q-error} is a common metric in database systems research for evaluating the accuracy of cardinality estimation models~\cite{cebench1,harmouch2017cardinality,iris}. It calculates the relative deviation of the predicted cardinality from the true cardinality for a given query. The Q-error for a single query is calculated as follows: 

% \begin{equation}
%     \textrm{Q-error} = \max \big( \frac{\hat{y}}{y}, \frac{y}{\hat{y}} \big) \in [1, +\infty),
% \end{equation}

% where $y$ and $\hat{y}$ represent true and predicted cardinalities, respectively. We particularly focus on P50 and P95 q-errors aggregated across all queries within a dataset to evaluate the models' accuracy in common and tail cases. 

\begin{figure}[tb]
\center
\includegraphics[width=0.98\textwidth]{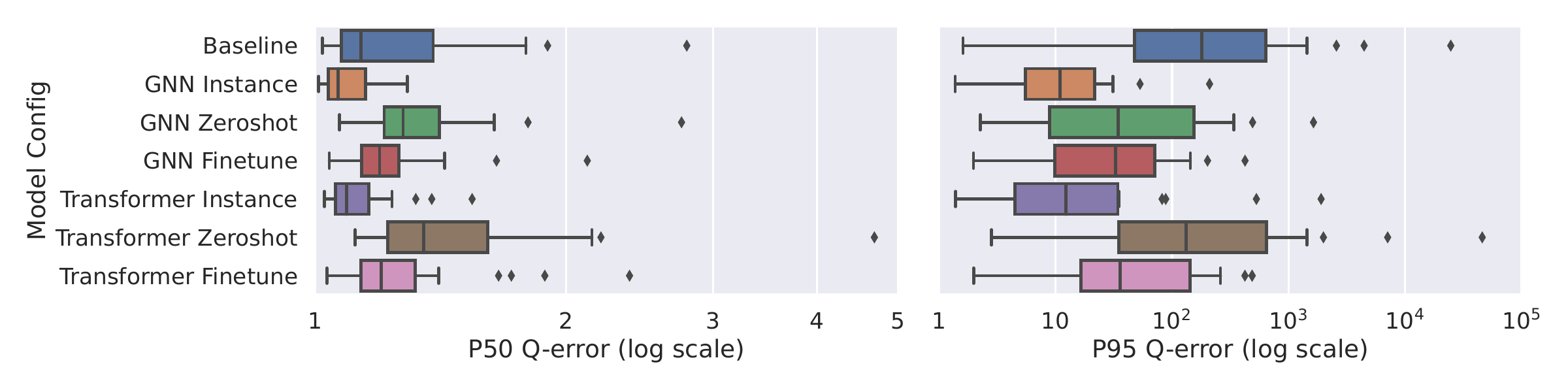}
\vspace{-4ex}
\caption{Box plots of P50 (left) and P95 (right) q-errors of different model configurations for queries on a single table, aggregated across 20 test datasets.}
\vspace{-0.2in}
\label{fig:accuracy_no_join}
\end{figure}

\begin{figure}[tb]
\center
\includegraphics[width=0.98\textwidth]{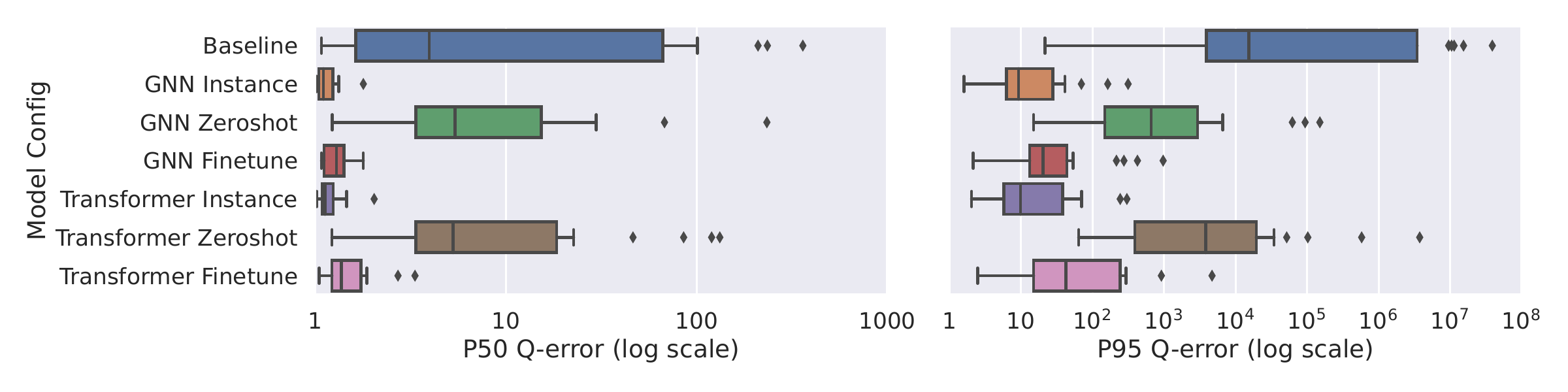}
\vspace{-4ex}
\caption{Box plots of P50 (left) and P95 (right) q-errors of different model configurations for queries contains a binary join, aggregated across 20 test datasets.}
\vspace{-0.2in}
\label{fig:accuracy_binary_join}
\end{figure}

% \vspace{-0.2in}

% Plots: https://colab.corp.google.com/drive/1qvVfpf0_QbdFe5oiCtGyYbmDTNpESWFV#scrollTo=SoxBPtPTY302
\subsection{Accuracy of Cardinality Prediction} % baseline, instance-4500, zeroshot-4500, finetune-500, P50 & P95
We evaluate the accuracy of cardinality prediction by comparing the baseline method, GNN models, and graph transformer models, each under different configurations as described in Section \ref{sec:experiments.model_configs}. We run a single experiment on each of the 20 test datasets in CardBench per model configuration. We then collect and aggregate the accuracy metrics across all experiments per model configuration. For training instance-based and zero-shot models, 4500 queries are randomly selected per dataset to form the training and validation set. Fine-tuning utilizes a smaller subset of 500 queries per dataset.

\subsubsection{Single Table Cardinality Prediction} 
Figure \ref{fig:accuracy_no_join} shows the box-and-whisker plots of P50 and P95 q-errors of different model configurations for queries on a single table, aggregated across 20 test datasets in CardBench. The baseline algorithm demonstrates strong median accuracy results in single-table cardinality prediction, with the majority of datasets achieving a q-error below $1.5$. 
%Estimating the cardinality of simple queries using pre-computed predicate selectivity, assuming independence for conjunctive and disjunctive predicates. 
However, accuracy degrades significantly at the tail due to the failure of both the uniform distribution and independence assumptions that the baseline approach makes, especially for more complex queries.

GNN-based models outperform transformer-based models in most cases, despite the latter's larger receptive field through attention mechanisms. This suggests that iterative message passing along query graph paths adequately captures the structure of SQL queries. Notably, instance-based models achieve the highest accuracy, with the GNN model demonstrating an average P50 q-error of $1.1$ and a P95 q-error of $24.08$, and the transformer model demonstrating an average P50 q-error of $1.15$ and a P95 q-error of $136.87$. It is achieved by implicitly learning dataset-specific joint-column data distributions, enabling tailored predictions. In contrast to instance-based models, zero-shot models struggle to generalize such knowledge to unseen datasets, resulting in a lower accuracy (especially at the tail). Fine-tuning pre-trained models, even with a modest 500 samples, significantly improves accuracy for both GNN and transformer models, highlighting the importance of incorporating table and column-specific distributions for accurate cardinality prediction.

\subsubsection{Binary Join Cardinality Prediction}
Cardinality prediction for queries with binary joins is much more challenging than single table because cross-table distribution and multi-column correlation, which are difficult to model, significantly impact join cardinality. Figure \ref{fig:accuracy_binary_join} shows the P50 and P95 q-errors for queries with binary joins. In contrast to single-table prediction, the baseline algorithm exhibits significant inaccuracy in binary join query CE, with a median q-error of $55.54$ and P95 q-error of $4.4 \times 10^6$, respectively. This appears to be a result of the oversimplified assumption that queries join two tables via single-column primary and foreign keys, without considering more complex cross-table relationships that are intractable using traditional heuristics. 

Learning-based models significantly outperform the baseline though, particularly at the 95th percentile. Instance-based models, which learn cross-table distributions within a dataset, achieve low median q-errors of 1.16 (GNN) and 1.20 (transformer). While their P95 q-errors are higher (37.55 for GNN, 43.96 for transformer), the results suggest that estimating join cardinality via learned distributions is feasible. However, out-of-distribution cardinality prediction for binary joins proves more challenging, as zero-shot models experience a dramatic increase in q-errors (up to 20x at the median, 5300x at P95). As with single-table results, even a small amount of data used for fine-tuning improves the accuracy of pre-trained models considerably.

\begin{figure}[tb]
\center
\includegraphics[width=0.98\textwidth]{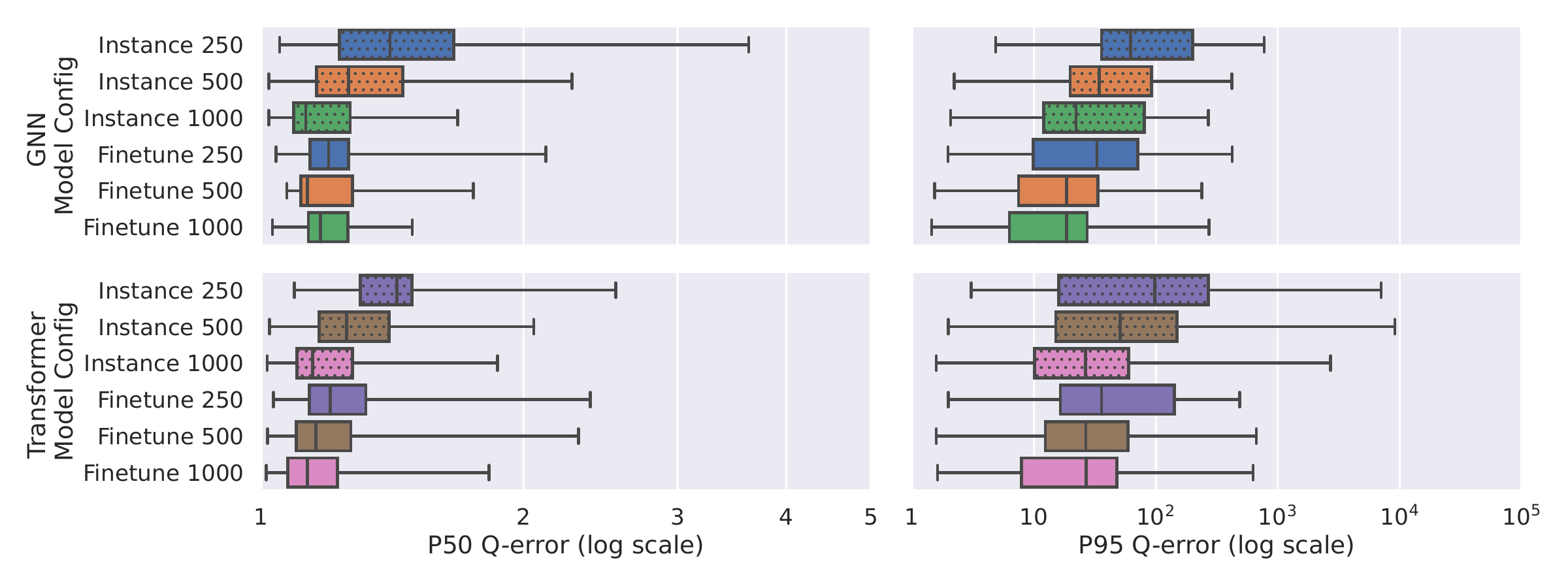}
\vspace{-2ex}
\caption{Box plots of P50 (left) and P95 (right) q-errors of instance-based vs. fine-tuned models of GNN (top) and Transformer (bottom) with varying training sample size for queries on a single table, aggregated across 20 test datasets.}
\vspace{-0.2in}
\label{fig:sample_size_no_join}
\end{figure}

\begin{figure}[tb]
\center
\includegraphics[width=0.98\textwidth]{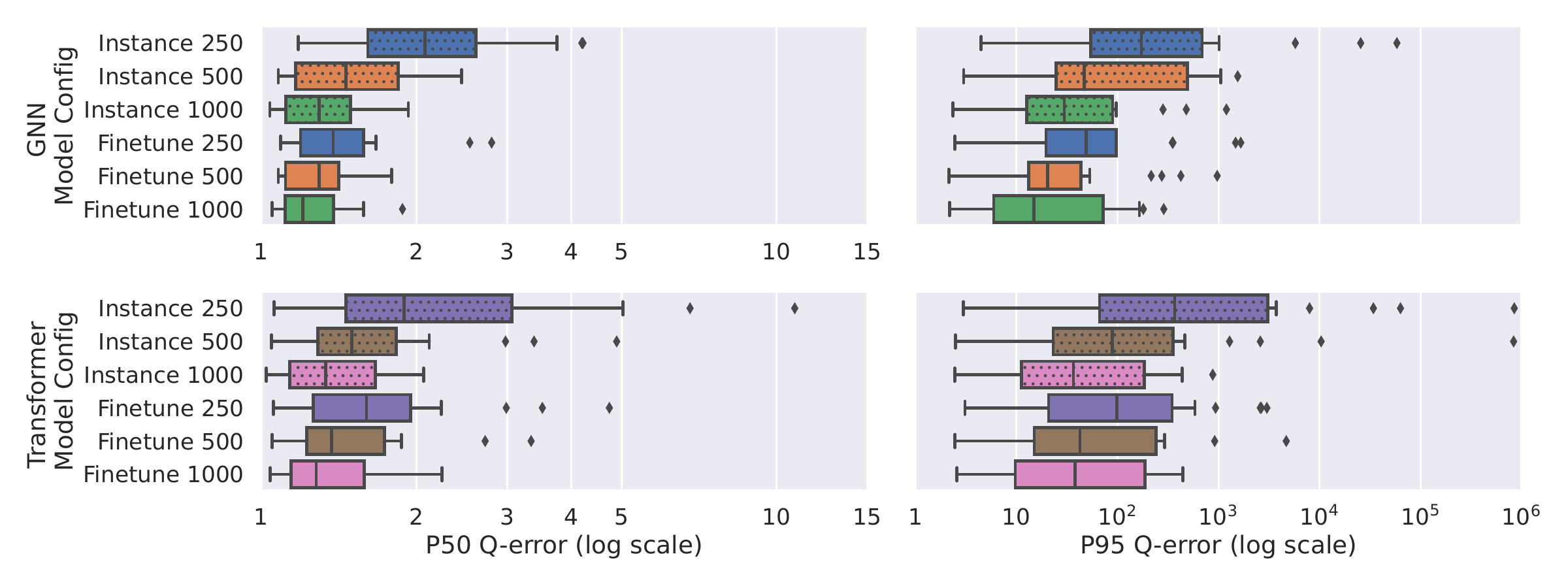}
\vspace{-2ex}
\caption{Box plots of P50 (left) and P95 (right) q-errors of instance-based vs. fine-tuned models of GNN (top) and Transformer (bottom) with varying training sample size for queries contains a binary join, aggregated across 20 test datasets.}
\vspace{-0.2in}
\label{fig:sample_size_binary_join}
\end{figure}

% \subsection{Sensitivity Study}
\subsection{Varying Sample Size} % instance-500 vs instance-4500 vs ft-500, P50 & P95 (if we have space, add instance-250&1000, ft-250&1000)
We study the impact of varying amounts of training data on model accuracy by considering three sample sizes: 250, 500, and 1000. For each sample size, we train an instance-based model or fine-tune a pre-trained model, and evaluate it on the same 500 samples as described in Section  \ref{sec:experiments.model_configs}. We repeat this process for both the GNN and transformer-based models. The results for single table and binary join cardinality predictions are summarized in figure \ref{fig:sample_size_no_join} and figure \ref{fig:sample_size_binary_join} respectively. We observe same patterns between GNN and transformer. The model accuracy improves as more training data is provided, for both instance-based and fine-tuned models.

Comparing instance-based training with fine-tuning, we find that the fine-tuned models are more sample-efficient. Fine-tuned models achieve similar median and tail Q-errors as instance-based models trained with twice the amount of data. This is evidence that fine-tuned models benefit from the generalization power of a pre-trained model, which has learned from a large number of data patterns over many datasets.
%Given limited training samples, there is more value in using them to fine-tune a pre-trained model as opposed to training a new model from scratch. When using only 500 samples on single table datasets, the average P50 and P95 Q-errors across 20 datasets are: 1.34 & 83 for the instanced-based GNN model, and 1.23 & 42 for the fine-tuned GNN model. 
%The improvement from fine-tuning with 500 samples is more prominent for cardinality prediction of binary joins, 
When using only 500 samples from each of the binary join datasets, the average P50 and P95 Q-errors across 20 datasets are: 1.57 \& 280 for the instanced-based GNN model, and 1.32 \& 120 for the fine-tuned GNN model. There is a significant improvement especially for the tail queries due to learnings transferred from more diverse data patterns seen by the pre-trained model. Nonetheless, the accuracy gap 
between instance-based and fine-tuned models reduces as the sample size increases. This suggests that fine-tuning on a pre-trained model is more advantageous when there is limited amount of training data available for a new dataset.

% \subsubsection{Number of Training Epochs} % 100 epochs vs 20 (Optional)

% Optional 
% \subsection{Performance} % training (instance-4500 vs. zeroshot-4500 vs finetune-500) and inference time (batch = 1,2,4,8,16,32, etc. ), GNN & Transformer
% Finally, we use training and inference time to measure the performance of cardinality prediction models. 
% \subsubsection{Training Time}
% \subsubsection{Inference Time}

\vspace{-0.15in}

%% file: contents/conclusions.tex
\section{Conclusion and Broader Impact} %TBD
\vspace{-0.1in}

In this paper, we release CardBench a benchmark for CE in relational databases. CardBench contains code to 1-) compute various data statistics, 2-) generate queries and execute them, and 3-) create the training data (the annotated querygraphs). We are also releasing two training datasets with queries over 20 datasets and their true cardinalities: one containing queries with single table queries and the other with binary joins. We trained both instance-based and zero-shot models, and observed that CE is a extremely challenging problem to learn in a zero-shot manner. However, when using pre-trained models with a small amount of fine-tuning, we can achieve accuracy similar to instance-based models, with much lower training overhead. 

With CardBench, we hope to provide a systematic mechanism to track progress, especially on recent directions such as pre-trained CE models.
Moreover, we think that CardBench is highly important to foster further research on CE models, from the ML and DB communities.
The training datasets were constructed with high overhead (7 cpu years of query runtime). Releasing such a training set lowers the barrier of entry for developing and training new CE models. Furthermore, we hope that the benchmark itself will be extended with new datasets and more complex queries using the tools and code we provide and thus push the boundaries of learned CE models further. 
%, or  generate new workloads with more complex queries, including multiple joins and aggregations.  

%We believe for CE models to be widely adapted in the industry, pre-trained models are critical and our initial results show the feasibility of such pre-trained CE models. More research is needed for pre-trained CE models  and our benchmark will be an invaluable resource to instigate further research in this direction.

%% file: contents/appendix.tex
\section{Appendix}

\subsection{Primary-Foreign Key Relationships} 
\label{sec:pkfk}
Primary foreign key (pk-fk) relationships are part of a database schema design~\cite{cow}. A pk-fk relationship is a logical connection between the columns of two tables semantically linking their information. In the real world, most of the times two tables are joined using the columns that participate in the pk-fk relationship. Naturally we want to use pk-fk relationships when creating synthetic workloads. A challenge we faced is that open source datasets do not usually specify pk-fk relationships, instead we discover pk-fk relationship between columns of different tables using the steps outlined below. One assumptions we make is that data is organized in a snowflake schema, this is the norm in practice and what is assumed by most database benchmarks (TPC-H \cite{tpch}, SSB \cite{ssb}, TPC-DS\cite{tpcds}, TPC-C\cite{tpcc}, JOB\cite{qoleis}).

\begin{enumerate}
  \item Identify candidate primary keys per table using a uniqueness threshold and choose one with the most unique values. A primary key should be unique for each row of a table~\cite{cow}, we relax this condition and require that for a column to be a primary key more than 90\% of rows must be unique values.
  \item Split tables of a dataset in fact and dimension tables\footnote{\url{https://en.wikipedia.org/wiki/Snowflake_schema}}. The criteria we use is a size threshold (fact tables should be a few times larger than dimension tables) as well as a check if a primary key exists for dimension tables.
  \item For each dimension table and pk column identify a fk column in the fact table(s). Out of all possible candidates we choose the pk-fk pair with the most join results, which we obtain by running a join query using the candidate pk-fk column pair as the join condition.
\end{enumerate}

\subsection{Graph Transformer} 
\begin{figure}[t!]
\centering
\includegraphics[width=0.4\textwidth]{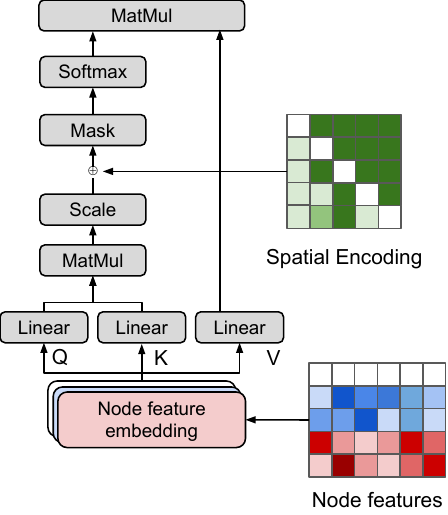}
\caption{An overview of the attention block of the graph transformer}
\label{fig:graph_transformer_overview}
\end{figure}

Figure \ref{fig:graph_transformer_overview} shows an overview of attention block of the graph transformer. 

%% file: arxiv_cardbench.bbl
\begin{thebibliography}{10}

\bibitem{tpcc}
T.~P.~P. Council.
\newblock {TPC-C benchmark}.
\newblock \url{https://www.tpc.org/tpcc/}, accessed 08/2024.

\bibitem{tpcds}
T.~P.~P. Council.
\newblock {TPC-DS benchmark}.
\newblock \url{https://www.tpc.org/tpcds/}, accessed 08/2024.

\bibitem{tpch}
T.~P.~P. Council.
\newblock {TPC-H benchmark}.
\newblock \url{https://www.tpc.org/tpch/}, accessed 08/2024.

\bibitem{cebench1}
Y.~Han, Z.~Wu, P.~Wu, R.~Zhu, J.~Yang, L.~W. Tan, K.~Zeng, G.~Cong, Y.~Qin,
  A.~Pfadler, Z.~Qian, J.~Zhou, J.~Li, and B.~Cui.
\newblock Cardinality estimation in {DBMS:} {A} comprehensive benchmark
  evaluation.
\newblock {\em Proc. {VLDB} Endow.}, 15(4):752--765, 2021.

\bibitem{cesurvey}
H.~Harmouch and F.~Naumann.
\newblock Cardinality estimation: An experimental survey.
\newblock {\em Proc. {VLDB} Endow.}, 11(4):499--512, 2017.

\bibitem{carsten}
B.~Hilprecht and C.~Binnig.
\newblock One model to rule them all: Towards zero-shot learning for databases.
\newblock In {\em 12th Conference on Innovative Data Systems Research, {CIDR}
  2022, Chaminade, CA, USA, January 9-12, 2022}. www.cidrdb.org, 2022.

\bibitem{deepdb}
B.~Hilprecht, A.~Schmidt, M.~Kulessa, A.~Molina, K.~Kersting, and C.~Binnig.
\newblock Deepdb: Learn from data, not from queries!
\newblock {\em Proc. {VLDB} Endow.}, 13(7):992--1005, 2020.

\bibitem{DBLP:conf/sigmod/KimJSHCC22}
K.~Kim, J.~Jung, I.~Seo, W.~Han, K.~Choi, and J.~Chong.
\newblock Learned cardinality estimation: An in-depth study.
\newblock In Z.~G. Ives, A.~Bonifati, and A.~E. Abbadi, editors, {\em {SIGMOD}
  '22: International Conference on Management of Data, Philadelphia, PA, USA,
  June 12 - 17, 2022}, pages 1214--1227. {ACM}, 2022.

\bibitem{mscn}
A.~Kipf, T.~Kipf, B.~Radke, V.~Leis, P.~A. Boncz, and A.~Kemper.
\newblock Learned cardinalities: Estimating correlated joins with deep
  learning.
\newblock In {\em 9th Biennial Conference on Innovative Data Systems Research,
  {CIDR} 2019, Asilomar, CA, USA, January 13-16, 2019, Online Proceedings}.
  www.cidrdb.org, 2019.

\bibitem{qoleis}
V.~Leis, A.~Gubichev, A.~Mirchev, P.~Boncz, A.~Kemper, and T.~Neumann.
\newblock How good are query optimizers, really?
\newblock {\em Proc. VLDB Endow.}, 9(3):204–215, nov 2015.

\bibitem{job}
V.~Leis, B.~Radke, A.~Gubichev, A.~Mirchev, P.~A. Boncz, A.~Kemper, and
  T.~Neumann.
\newblock Query optimization through the looking glass, and what we found
  running the join order benchmark.
\newblock {\em {VLDB} J.}, 27(5):643--668, 2018.

\bibitem{iris}
Y.~Lu, S.~Kandula, A.~C. K{\"{o}}nig, and S.~Chaudhuri.
\newblock Pre-training summarization models of structured datasets for
  cardinality estimation.
\newblock {\em Proc. {VLDB} Endow.}, 15(3):414--426, 2021.

\bibitem{cebench2}
P.~Negi, R.~Marcus, A.~Kipf, H.~Mao, N.~Tatbul, T.~Kraska, and M.~Alizadeh.
\newblock Flow-loss: Learning cardinality estimates that matter.
\newblock {\em Proc. {VLDB} Endow.}, 14(11):2019--2032, 2021.

\bibitem{DBLP:conf/icde/NegiMMTKA20}
P.~Negi, R.~Marcus, H.~Mao, N.~Tatbul, T.~Kraska, and M.~Alizadeh.
\newblock Cost-guided cardinality estimation: Focus where it matters.
\newblock In {\em 36th {IEEE} International Conference on Data Engineering
  Workshops, {ICDE} Workshops 2020, Dallas, TX, USA, April 20-24, 2020}, pages
  154--157. {IEEE}, 2020.

\bibitem{DBLP:journals/pvldb/NegiWKTMMKA23}
P.~Negi, Z.~Wu, A.~Kipf, N.~Tatbul, R.~Marcus, S.~Madden, T.~Kraska, and
  M.~Alizadeh.
\newblock Robust query driven cardinality estimation under changing workloads.
\newblock {\em Proc. {VLDB} Endow.}, 16(6):1520--1533, 2023.

\bibitem{ssb}
P.~O’Neil, E.~O’Neil, X.~Chen, and S.~Revilak.
\newblock The star schema benchmark and augmented fact table indexing.
\newblock In {\em Performance Evaluation and Benchmarking: First TPC Technology
  Conference, TPCTC 2009, Lyon, France, August 24-28, 2009, Revised Selected
  Papers 1}, pages 237--252. Springer, 2009.

\bibitem{cow}
R.~Ramakrishnan and J.~Gehrke.
\newblock {\em Database management systems}.
\newblock McGraw-Hill, Inc., 2002.

\bibitem{accesspath}
P.~G. Selinger, M.~M. Astrahan, D.~D. Chamberlin, R.~A. Lorie, and T.~G. Price.
\newblock Access path selection in a relational database management system.
\newblock In {\em Proceedings of the 1979 ACM SIGMOD International Conference
  on Management of Data}, SIGMOD '79, page 23–34, New York, NY, USA, 1979.
  Association for Computing Machinery.

\bibitem{DBLP:journals/pvldb/SunZSLT21}
J.~Sun, J.~Zhang, Z.~Sun, G.~Li, and N.~Tang.
\newblock Learned cardinality estimation: {A} design space exploration and {A}
  comparative evaluation.
\newblock {\em Proc. {VLDB} Endow.}, 15(1):85--97, 2021.

\bibitem{DBLP:journals/pvldb/WangCLL21}
J.~Wang, C.~Chai, J.~Liu, and G.~Li.
\newblock {FACE:} {A} normalizing flow based cardinality estimator.
\newblock {\em Proc. {VLDB} Endow.}, 15(1):72--84, 2021.

\bibitem{balsa}
Z.~Yang, W.-L. Chiang, S.~Luan, G.~Mittal, M.~Luo, and I.~Stoica.
\newblock Balsa: Learning a query optimizer without expert demonstrations.
\newblock In {\em Proceedings of the 2022 International Conference on
  Management of Data}, SIGMOD '22, page 931–944, New York, NY, USA, 2022.
  Association for Computing Machinery.

\bibitem{neurocard}
Z.~Yang, A.~Kamsetty, S.~Luan, E.~Liang, Y.~Duan, X.~Chen, and I.~Stoica.
\newblock Neurocard: One cardinality estimator for all tables.
\newblock {\em Proc. {VLDB} Endow.}, 14(1):61--73, 2020.

\bibitem{ying2021transformers}
C.~Ying, T.~Cai, S.~Luo, S.~Zheng, G.~Ke, D.~He, Y.~Shen, and T.-Y. Liu.
\newblock Do transformers really perform badly for graph representation?
\newblock {\em Advances in neural information processing systems},
  34:28877--28888, 2021.

\bibitem{yun2019graph}
S.~Yun, M.~Jeong, R.~Kim, J.~Kang, and H.~J. Kim.
\newblock Graph transformer networks.
\newblock {\em Advances in neural information processing systems}, 32, 2019.

\bibitem{DBLP:journals/pvldb/ZhuWHZPQZC21}
R.~Zhu, Z.~Wu, Y.~Han, K.~Zeng, A.~Pfadler, Z.~Qian, J.~Zhou, and B.~Cui.
\newblock {FLAT:} fast, lightweight and accurate method for cardinality
  estimation.
\newblock {\em Proc. {VLDB} Endow.}, 14(9):1489--1502, 2021.

\end{thebibliography}
